\documentclass[english,journal=aamick,manuscript=article]{achemso}
\usepackage[T1]{fontenc}
\usepackage[latin9]{inputenc}
\usepackage{textcomp}
\usepackage{amsmath}
\usepackage{amsthm}
\usepackage{amssymb}
\usepackage{graphicx}

\makeatletter

\title{Localized polarons and conductive charge carriers: understanding
CaCu$_{3}$Ti$_{4}$O$_{12}$ over a broad temperature range}

\author{Laijun Liu}

\affiliation{Current address: College of Materials Science and Engineering, Guilin
University of Technology, Guilin, 541004, China}

\author{Shaokai Ren}

\affiliation{Current address: College of Materials Science and Engineering, Guilin
University of Technology, Guilin, 541004, China}

\author{Jia Liu}

\affiliation{Current address: State Key Laboratory for Mechanical Behavior of
Materials, School of Materials Science and Engineering, Xi'an Jiaotong
University, Xi'an 710049 China}

\author{Feifei Han}

\affiliation{Current address: College of Materials Science and Engineering, Guilin
University of Technology, Guilin, 541004, China}

\author{Jie Zhang}

\affiliation{Current address: Electronic Materials Research Laboratory-Key Laboratory
of the Ministry of Education and International Center for Dielectric
Research, Xi'an Jiaotong University, Xi'an 710049, China}

\author{Biaolin Peng}

\affiliation{Current address: School of Physical Science \& Technology and Guangxi
Key Laboratory for Relativistic Astrophysics, Guangxi University,
Nanning 530004, China}

\author{Dawei Wang}

\email{dawei.wang@mail.xjtu.edu.cn }

\affiliation{Current address: School of Microelectronics \& State Key Laboratory
for Mechanical Behavior of Materials, Xi'an Jiaotong University, Xi'an
710049, China}

\author{Alexei A. Bokov}

\affiliation{Current address: Department of Chemistry and 4D LABS, Simon Fraser
University, Burnaby, British Columbia, V5A 1A6, Canada}

\author{Zuo-Guang Ye}

\affiliation{Current address: Department of Chemistry and 4D LABS, Simon Fraser
University, Burnaby, British Columbia, V5A 1A6, Canada}


\makeatother

\usepackage{babel}
\begin{document}
\begin{abstract}
CaCu$_{3}$Ti$_{4}$O$_{12}$ (CCTO) has a large dielectric permittivity
that is independent of the probing frequency near the room temperature,
which complicated due to the existence of several dynamic processes.
Here, we consider the combined effects of localized charge carriers
(polarons) and thermally activated charge carriers using a recently
proposed statistical model to fit and understand the permittivity
of CCTO measured at different frequencies over the whole temperature
range accessible by our experiments. We found that the small permittivity
at the lowest temperature is related to polaron frozen, while at higher
temperatures the rapid increase is associated with the thermal excitation
of polarons inducing the Maxwell-Wagner effect, and the final increase
of the permittivity is attributed to the thermally activated conductivity.
Such analysis enables us to separate the contributions from localized
polarons and conductive charge carriers and quantify their activation
energies.
\end{abstract}

\section{Introduction}

Materials with high dielectric permittivity have attracted much attention
due to their numerous technological applications. CaCu$_{3}$Ti$_{4}$O$_{12}$
(CCTO) has a cubic perovskite structure and a giant relative permittivity
of $10^{4}\sim10^{6}$ near the room temperature. While the dielectric
permittivity of CCTO exhibits a very small temperature dependence
between 100\,K and 600\,K, it drops rapidly to a value of $\sim100$
below 100\,K.\cite{Subramanian,Ramirez,Homes,Lin,Si,Felix,Schmidt,Han,Deng,Deng10}
This overall behavior of CCTO is very different from either relaxors
or normal ferroelectrics. It is established that the origin of such
high permittivity is an extrinsic effect, which in polycrystalline
CCTO is modeled with internal barrier layer capacitor (IBLC), consisting
of semiconducting grains separated by thin insulating grain boundaries
leading to high dielectric permittivity. Furthermore, the interface
effect from electrodes or domain walls is suggested to also contribute
to the high permittivity especially in single crystals. 

In addition, such remarkable dielectric properties also strongly depend
on the probing frequency, which approximately follows the Arrhenius
behavior. Similar phenomenon was also found in many manganites, cuprates,
and nickelates, such as La$_{1-x}$Ca$_{x}$MnO$_{3}$,\cite{Neupane}
Pr$_{0.7}$Ca$_{0.3}$MnO$_{3}$,\cite{Freitas} Tb/EuMnO$_{3}$,\cite{Wang,Yang,Deng1,Deng2}
CuTa$_{2}$O$_{6}$,\cite{LiG} LaCuLiO$_{4}$,\cite{ParkT} LaSrNiO$_{4}$,
\cite{Rivas} Li/Ti-doped NiO\cite{Wu,LiY} and Ba(Fe$_{0.5}$Nb$_{0.5}$)O$_{3}$.\cite{Raevski,Ke,Huang,Lius,SunX}
Therefore, a better understanding of CCTO will also help to explain
the dielectric properties of a large group of materials. For these
materials, their low-temperature dielectric relaxation has been attributed
to hopping of polarons, which are localized charge carriers interacting
with phonons \cite{WangCC}, between lattice sites with a characteristic
timescale. The step-function-like decrease suggests a freezing temperature
in the relaxation behavior following a glass-like process. 

However, important questions regarding this type of materials remain
unanswered. For instance, while it is known that the polaronic relaxation
usually involves either a variable range hopping (VRH) or a nearest-neighbor
hopping conduction process,\cite{ZhangLL,Tselev} an estimation of
the activation energy, which is a key parameter for such process,
is still missing. Moreover, at higher temperatures, the permittivity
of such materials also include the contribution from thermally activated
conducting electrons. The non-localized conductivity is not a pure
\emph{dc} conductivity, that is, high dielectric loss is accompanied
by the increase of dielectric permittivity at low frequencies. It
is unclear how important such contribution is to the total permittivity,
which contains the effects of both thermally activated polarons and
conductive charge carriers. As a matter of fact, for the permittivity
of such materials, a complete description of their temperature dependence
is unavailable over the temperature range accessible to experiments.
It is worth noting that the Maxwell-Wagner model alone is not enough
to understand the permittivity over the whole temperature range (see
Eq. \eqref{eq:total-permittivity} below). Moreover, while the system
may be modeled with a parallel RC equivalent circuit and results in
the Arrhenius equation,\cite{Valdez-Nava} the physics underlying
such phenomenon needs further understanding to establish connections
between the RC circuits and charge carriers in the system. 

In this paper, we will answer the aforementioned questions using a
statistical approach, which is based on macroscopic and phenomenological
considerations of both thermally activated polarons and conductive
charge carriers in CCTO. With this approach, we are able to propose
an explicit formula to fit the permittivity over the whole temperature
range accessible by experiments, which in turn enables us to separate
the contributions from polaron and conductive charge carriers. Such
separation finally allows us to estimate the activation energies of
polarons, as well as conductive charge carriers. Since the analysis
separates these two effects and offer some details regarding them,
it will enhance our understanding of CCTO and similar materials, and
provide useful clues to design or improve this type of materials.

\section{Experimental Section}

CaCu$_{3}$Ti$_{4}$O$_{12}$ (CCTO) powder was prepared by a molten
salt method.\cite{LiuLJ} The obtained CCTO powder was pressed into
pellets of 15mm in diameter and $\sim1$\,mm in thickness. The pellets
were sintered at 1060\,\textcelsius{} in the air for 30\,h. X-ray
diffraction shows that the powder is of pure cubic perovskite phase.
Both sides of the samples were first polished and then brushed with
silver conductive paste, which is followed by a heat treatment at
550\,\textcelsius{} for 30\,min. Dielectric measurements were performed
with an applied voltage of 500\,mV using an Agilent 4294A impedance
analyzer over the frequency range of 100\,Hz to 1\,MHz and over
the temperature range from 90\,K to 500\,K.

\section{Results and discussion}

\begin{figure}
\begin{centering}
\includegraphics[width=10cm]{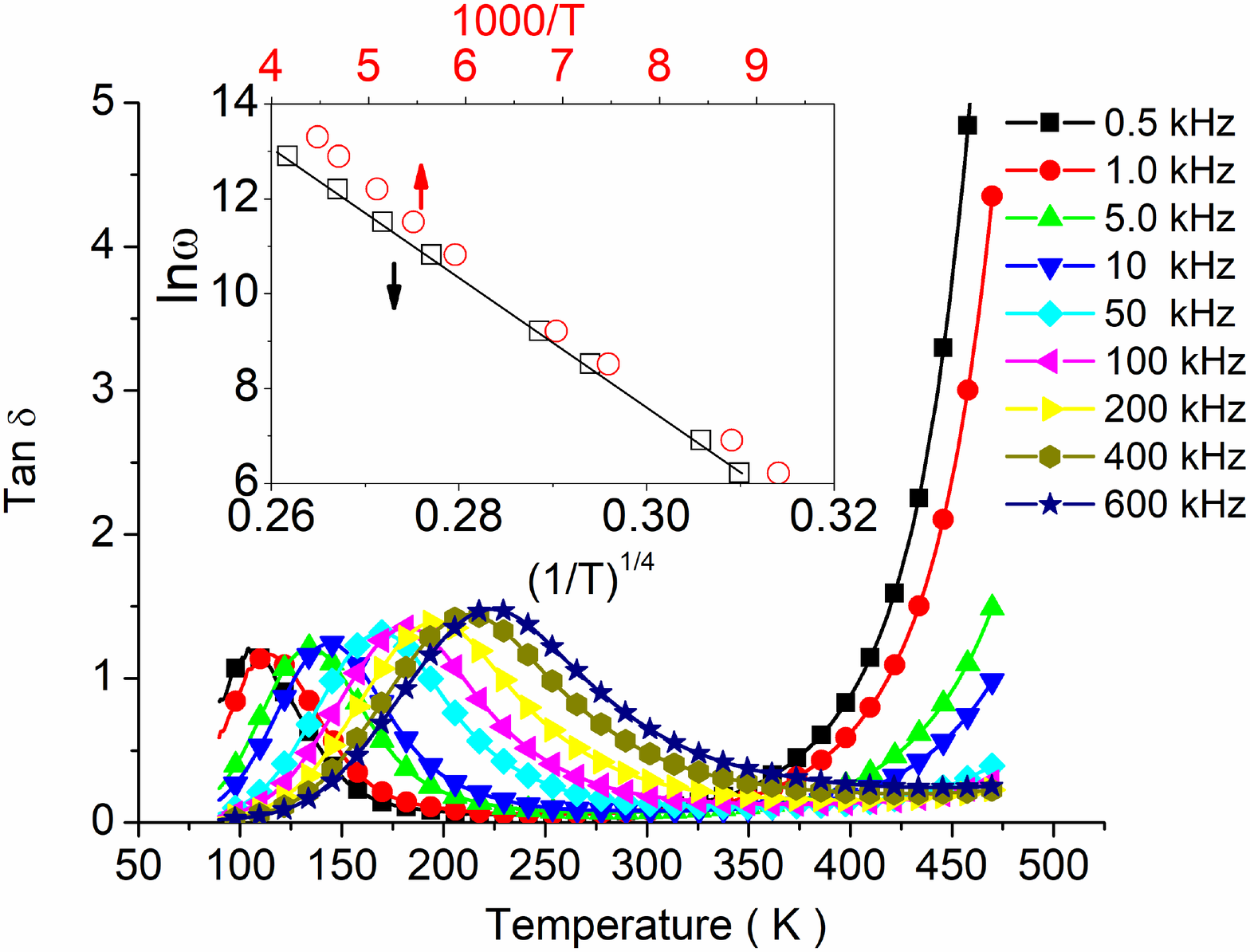}
\par\end{centering}
\caption{Temperature dependence of $\textrm{tan}\delta$ of the permittivity
of CCTO at various frequencies. Inset: temperature-dependent relaxation
frequency $\omega$ scaled for thermally activated nearest-neighbor
hopping of charges (open circles, top and left axes) and VRH (open
squares, bottom and left axes). The solid line is the fitting curve
of the experimental data (open squares) according to Eq. \ref{eq:VRH}.\label{fig1:Temperature_vs_dielectric_loss}}
\end{figure}

The temperature dependence of the loss factor of the CCTO sample is
shown in Fig. \ref{fig1:Temperature_vs_dielectric_loss}. Below 300\,K,
the loss peaks shift to high temperature with the probing frequency,
which shows a Debye-like relaxation. At high temperatures, the loss
factor increases rapidly with temperature, suggesting a thermally
activated process. 

\subsection{Dielectric relaxation}

To understand the relaxation at low temperatures, we first analyze
the behavior of the dielectric dissipation. The inset in Fig. \ref{fig1:Temperature_vs_dielectric_loss}
shows the temperature dependence of the relaxation frequency of CCTO,
$\ln\omega$ vs $1/T^{1/4}$ (open squares), where $\omega$ is the
position of the loss peak in the $\textrm{tan}\delta$ versus $\ln\omega$
plots. We can clearly see that there is a very good linear relation
between $\textrm{ln}\mathit{\omega}$ and $1/T^{1/4}$.

On the other hand, if we plot $\textrm{ln}\omega$ as a function of
$1000/T$ (open circles), the result shows an approximate Arrhenius
relation. According to this relation, the relaxation frequency at
the infinite temperature and the activation energy are found to be
$3.59\times10^{8}$\,Hz and 130\,meV, respectively. Theses values
are in good agreement with that of perovskite materials,\cite{LiuLJ2}
associated with the localization process of charge carriers. However,
it is notable that a distinct deviation from the Arrhenius relation
exists above $\sim200$\,K. Such a deviation has been found in polaron
related relaxations of certain materials, such as CCTO \cite{WangCC,ZhangLL,Bidault}
and Sr$_{0.998}$Ca$_{0.002}$TiO$_{3}$.\cite{Mott} The reason for
the deviation from the Arrhenius law is likely the transition from
a grain boundary-limited to bulk-limited conduction, consistent with
the widely held \textquotedbl barrier layer model\textquotedbl .\cite{Krohns,Chung} 

Therefore, cccording to our results and calculations, Mott's variable-range-hopping
(VRH) model,\cite{Kastner} i.e., 

\begin{equation}
f=f_{1}\textrm{exp}\left[-\left(T_{1}/T\right)^{1/4}\right]\label{eq:VRH}
\end{equation}
can be used to better fit the relaxation frequency, where $f_{1}$
and $T_{1}$ are two constants. The solid line in the inset of Fig.
\ref{fig1:Temperature_vs_dielectric_loss} is the fitting result of
our experimental results according to Eq. \ref{eq:VRH}. The values
of $f_{1}$ and $T_{1}$ are determined to be $1.68\times10^{21}$\,Hz
and $3.56\times10^{8}$\,K, respectively. The value of $T_{1}$ of
CCTO is similar to those of Li-doped La$_{2}$CuO$_{4}$ \cite{Ang}
and Cu-doped BaTiO$_{3}$ \cite{Karmakar} while $f_{1}$ is much
higher. According to the IBLC model, the relaxation frequency is related
to \emph{dc} conductivity and grain-boundary capacitance of CCTO \cite{ZhangLL}.
As a matter of fact, is was shown in Ref. \cite{ZhangLL} that $f_{1}$
has an approximate linear relation with the \emph{dc} conductivity
of the material. Therefore, $f_{1}$ does not represent the hopping
frequency of polarons.

Since the VRH mechanism describes the low-temperature dielectric relaxation
of CCTO ceramics well, it supports the idea that three-dimensional
disorder effects dominate the relaxation behavior of CCTO's semiconducting
phases. At this phase, the kinetic energy (due to the thermal excitation)
is insufficient to excite charge carrier across the Coulomb gap, therefore
Eq. \ref{eq:VRH} is mostly due to the hopping of charge carriers
within small regions. Figure \ref{fig1:Temperature_vs_dielectric_loss}
thus provides strong evidence for the existence of (hopping) polarons
in CCTO.

\subsection{Permittivity}

\begin{figure}
\begin{centering}
\includegraphics[width=12cm]{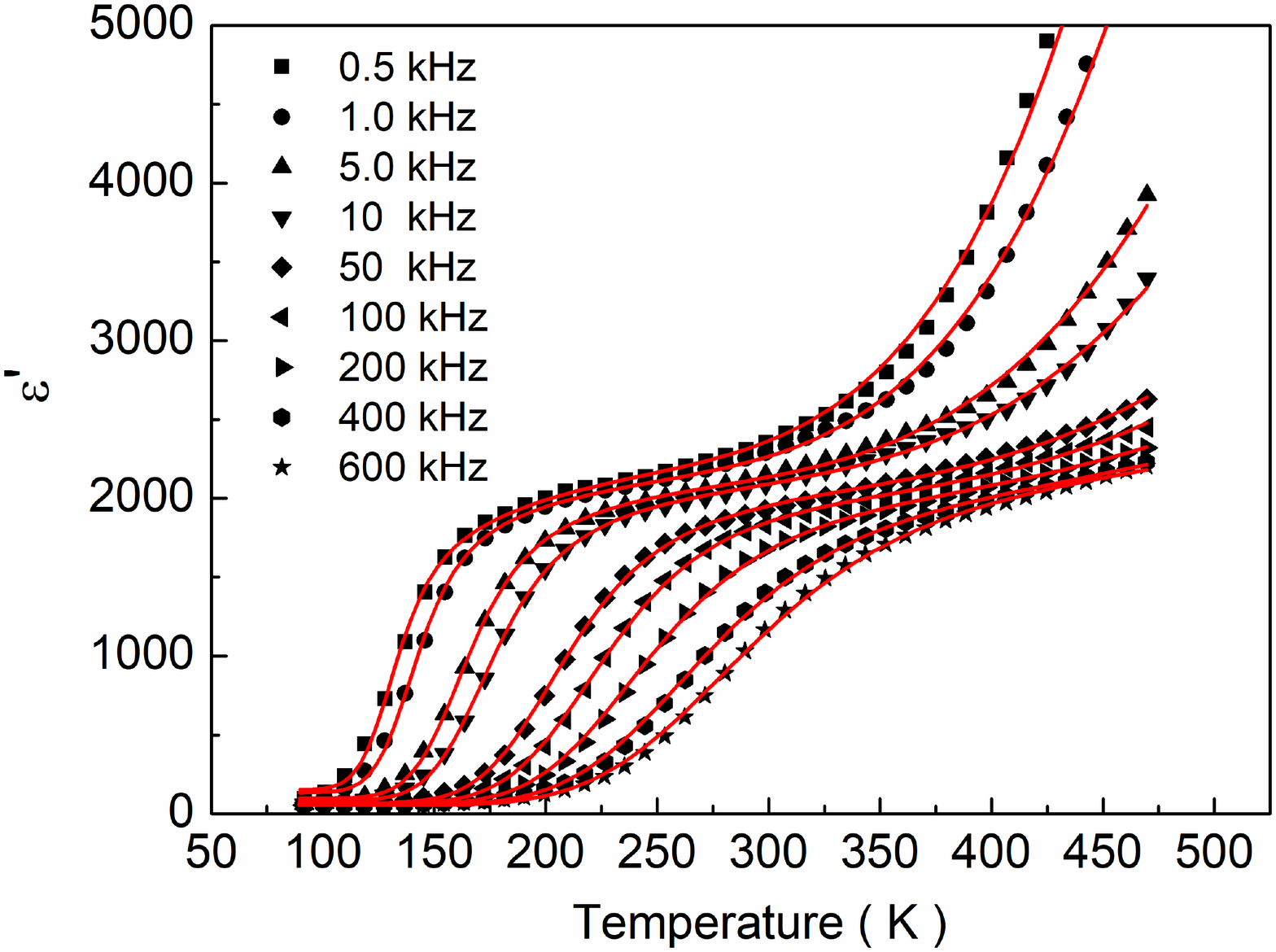}
\par\end{centering}
\caption{Temperature dependence of the real dielectric permittivity of the
CCTO with different frequencies. The solid lines are the fitting results
according to the Eq (4).\label{fig2:Temperature_vs_ Permittivity}}
\end{figure}
The temperature dependence of CCTO's permittivity measured at different
frequencies is shown in Fig. \ref{fig2:Temperature_vs_ Permittivity}.
The rapid increase in the low-temperature region is intimately related
to the polaron hopping. The curves in Fig. \ref{fig2:Temperature_vs_ Permittivity}
have four stages: (i) a plateau region at the lowest temperatures
(the exact temperature depends on the probing frequency; at 1\,kHz,
$T<90$\,K), where polarons are frozen and the permittivity is below
100. (ii) As the temperature increases, the dielectric permittivity
has a rapid rise (at 1 kHz, $90<T<150$\,K), which is related to
the thermal excitation of polarons inside grains. The polaron hopping
leads to semiconducting grains, where charge carriers are able to
move freely (but not beyond the grain boundaries). Consequently, such
grains separated by thin insulating grain boundaries result in the
Maxwell-Wagner effect, which significantly enhances the dielectric
permittivity. (iii) A plateau region exists at even higher temperatures
(at 1 kHz, $150<T<350$\,K). (iv) Another rapid increase of dielectric
permittivity at the highest temperatures, which can be attributed
to the thermally activated conductivity over the bulk. 

Looking carefully at these results, we can see some similarity to
the temperature dependent permittivity of relaxors,\cite{Liuj} where
it was found that contributions from different origins have varying
weight over the whole temperature range. Interestingly, we find that
the statistical approach adopted in Ref. \cite{Liuj} can also be
used to understand the dielectric response of polarons in CCTO as
well. 

\subsection{Fitting model}

The Maxwell-Wagner relaxation model can be simplified as two alternating
slabs having different dielectric permittivity ($\varepsilon_{\textrm{g}}$
and $\varepsilon_{\textrm{gb}}$ for grains and grain boundaries,
respectively), conductivity ($\sigma_{\textrm{g}}$ and $\sigma_{\textrm{gb}}$
for grains and grain boundaries, respectively) and widths ($l$ and
$d$ corresponding to grains and grain boundaries, respectively and
$L=l+d$), one can approximately obtain the effective permittivity
as \cite{Raevski2003a,Raevski2003b}

\begin{align}
\varepsilon^{\ast}= & \frac{L}{l/\left(\varepsilon_{\textrm{g}}-i\sigma_{\textrm{g}}/\omega\right)+d/\left(\varepsilon_{\textrm{gb}}-i\sigma_{\textrm{gb}}/\omega\right)}\nonumber \\
= & \varepsilon_{\infty}+\frac{\varepsilon_{0}-\varepsilon_{\infty}}{1+i\omega\tau}-i\frac{\sigma\left(\omega\right)}{\omega},\label{eq:permittivity-with-grains-and-boundaries}
\end{align}
where $\varepsilon_{\infty}=L/\left(l/\varepsilon_{\textrm{g}}+d/\varepsilon_{\textrm{gb}}\right)$,
$\varepsilon_{0}=L\left(\sigma_{\textrm{gb}}\varepsilon_{\textrm{g}}+\sigma_{g}\varepsilon_{\textrm{gb}}\right)/\left(l\sigma_{\textrm{gb}}+d\sigma_{\textrm{g}}\right)$,
$\tau=\left(l\varepsilon_{\textrm{gb}}+d\varepsilon_{\textrm{g}}\right)/\left(l\sigma_{\textrm{gb}}+d\sigma_{\textrm{g}}\right)$,
and $\sigma\left(\omega\right)=\sigma_{\textrm{g}}\sigma_{\textrm{gb}}/\left[\left(l\sigma_{\textrm{gb}}+d\sigma_{\textrm{g}}\right)\left(1+i\omega\tau\right)\right]$.
The dielectric permittivity shown in Eq. \eqref{eq:permittivity-with-grains-and-boundaries}
consists of three contributions. The first is a constant determined
by the permittivity of both the grain and its boundary. The second
contribution is of Debye type with the relaxation time determined
by the conductivity and permittivities. The third term describes the
contribution from the conductivity of grains and grain boundaries.
We can see that the relaxation time $\tau$ can strongly depends on
the temperature if the conductivity of the system can change significantly
with temperature.

The phenomenological model we propose is based on the Debye relaxation
of polarons and the \emph{ac} conductivity of conductive charge carriers.
In this model, individual polarons are categorized into two groups.
Our assumption is that the polarons need to be thermally excited to
overcome a local energy minimum before they can have large contributions
to the grain conductivity $\sigma_{\textrm{g}}$ and permittivity
$\varepsilon_{\textrm{g}}$. When the temperature is sufficiently
low, most of the polarons charge can only move around its equilibrium
position, their contribution to grain conductivity and permittivity
is small and can be taken as an insulator, where the permittivity
can be taken as a constant. As the temperature increases, more and
more of them can move on a larger spacial scale (but still bound by
grain boundaries), jumping from one energy minimum to others, with
their contribution to the conductivity and permittivity becoming much
larger.

Similar to our statistical model for relaxors,\cite{Liuj} here we
employ the Maxwell-Boltzmann distribution to estimate the number of
active polarons relative to the inactive ones, where we need to introduce
a potential well of average depth, $E_{b}$, to account for the constraint
on the polarons. One important reason we can apply the model developed
for relaxor is because polarons may be taken as individual particles
as the dipoles in relaxors. Practically, the number of polarons with
a kinetic energy exceeding the potential well $\left[N_{1}\left(E_{b},T\right)\right]$
is given by

\begin{equation}
N_{1}\left(E_{b},T\right)=N\sqrt{\frac{4}{\pi}}\sqrt{\frac{E_{b}}{k_{B}T}}\textrm{exp}\left(-\frac{E_{b}}{k_{B}T}\right)+N\textrm{erfc}\sqrt{\frac{E_{b}}{k_{B}T}},
\end{equation}
where $N$ is the total number of polarons in the system, $k_{B}$
is the Boltzmann constant, $T$ is the temperature (in Kelvin), and
erfc is the complementary error function. The total dielectric permittivity
is then given by

\begin{equation}
\varepsilon\left(T,\omega\right)=\varepsilon_{1}\left(T,\omega\right)P_{1}\left(E_{b},T\right)+\varepsilon_{2}\left(T,\omega\right)P_{2}\left(E_{b},T\right),\label{eq:permittivity}
\end{equation}
where $\varepsilon_{1}\left(T,\omega\right)$ and $\varepsilon_{2}\left(T,\omega\right)$
describe the dielectric responses from the aforementioned two polaron
groups, $\omega$ is the probing frequency, and $P_{1}\left(E_{b},T\right)=N_{1}\left(E_{b},T\right)/N$,
$P_{2}\left(E_{b},T\right)=1-P_{1}\left(E_{b},T\right)$ account for
the proportion of polarons in each group. We note that, according
to Eq. \eqref{eq:permittivity-with-grains-and-boundaries}, the permittivity
in Eq. \ref{eq:permittivity} result from not only the intrinsic polarization
of polarons, but also the grain conductivity associated with the distribution
of the thermally activated polarons, which is often described by the
Maxwell-Wagner effect. 

It has been found that the temperature where the dielectric permittivity
is almost a step function strongly depends on the probing frequency,
which approximately follows an Arrhenius behavior, therefore the Debye
relaxation, i.e., $\varepsilon\sim1/\left(1+\omega^{2}\tau_{0}^{2}\right)$
\cite{Kasap} was chosen, where $\tau_{0}$ is the temperature-dependent
relaxation time. For a thermally activated process, it follows the
Arrhenius law. Consequently, the permittivity will be $\varepsilon\sim1/\left[1+A^{2}\omega^{2}\textrm{exp}\left(2E_{a}/T\right)\right]$,
which is discussed by Jonscher \cite{Jonscher} and follows the approach
shown in Ref. \cite{Liuj}. 

In addition to the above analysis, we also need to consider the high-temperature
dielectric response induced by the conductivity due to the thermally
activated non-localized conductive charge carriers \cite{Maglione2016},
which is accounted for by a new term (the last term in the equation
below)

\begin{equation}
\varepsilon\left(T\right)=\frac{\varepsilon_{1}}{1+\omega^{2}\tau_{0}^{2}\textrm{exp}\left(-\theta/T\right)}P_{1}\left(E_{b},T\right)+\varepsilon_{2}P_{2}\left(E_{b},T\right)+\frac{\sigma\textrm{exp}\left(-E_{\textrm{con}}/k_{B}T\right)}{\varepsilon_{0}\omega},\label{eq:total-permittivity}
\end{equation}
where $\varepsilon_{1}$, $\varepsilon_{2}$, $\tau_{0}$ and $\theta$
are constants at a given frequency $\omega$, $\varepsilon_{0}$ is
the vacuum permittivity, $\sigma$ is the thermal activated conductivity,
and $E_{\textrm{con}}$ is the conductivity activation energy for
conductive charge carries' migration and transport. 

The above Eq. (4) has clear physical meanings: (i) The first and second
terms are the same as in Ref. \cite{Liuj} except that the analysis
applies to polarons; (ii) The third term or the RHS is the contribution
of conductive charge carriers, which is often associated with a thermal
activation process.

\subsection{Analysis using the fitting model}

Having shown the explicit formula to describe the dielectric permittivity
versus temperature, we now use Eq. (4) to fit $\varepsilon\left(T\right)$
of CCTO at different frequencies and show the results in Fig. \ref{fig2:Temperature_vs_ Permittivity}
(solid line). The fitting curves are in very good agreement with experimental
results. For CCTO, the $E_{b}$ of the polarons exhibits very little
dependence on the probing frequency. Therefore, we find $E_{b}=0.19$\,eV
with a simple averaging procedure and use this value for all the fittings
at different frequencies.

The activation energy for the conducting charge carriers $E_{\textrm{con}}$
usually is more related to the composition rather than the probing
frequency. From our fitting results, $E_{\textrm{con}}$=0.25\,eV
is also obtained with a simple averaging procedure. The $\theta$
value is 2100\,K, which only slightly changes with the probing frequency,
while $\textrm{ln}\tau_{0}$ is inversely proportional to the probing
frequency $\omega$. The good fitting in Fig. \ref{fig2:Temperature_vs_ Permittivity}
shows the importance of polaronic conduction in the grains, whose
relaxation was likely induced by the charge accumulation at the grain
boundaries (because grain boundaries are less conductive), which reveals
some connections between polarons and the IBCL model. We also note
that the fitting parameters (e.g., $E_{\textrm{con}}$) depend on
samples, especially their grain sizes and composition as discussed
in Ref. \cite{LiuLJ3}. 

\begin{figure*}
\noindent \begin{centering}
\includegraphics[width=14cm]{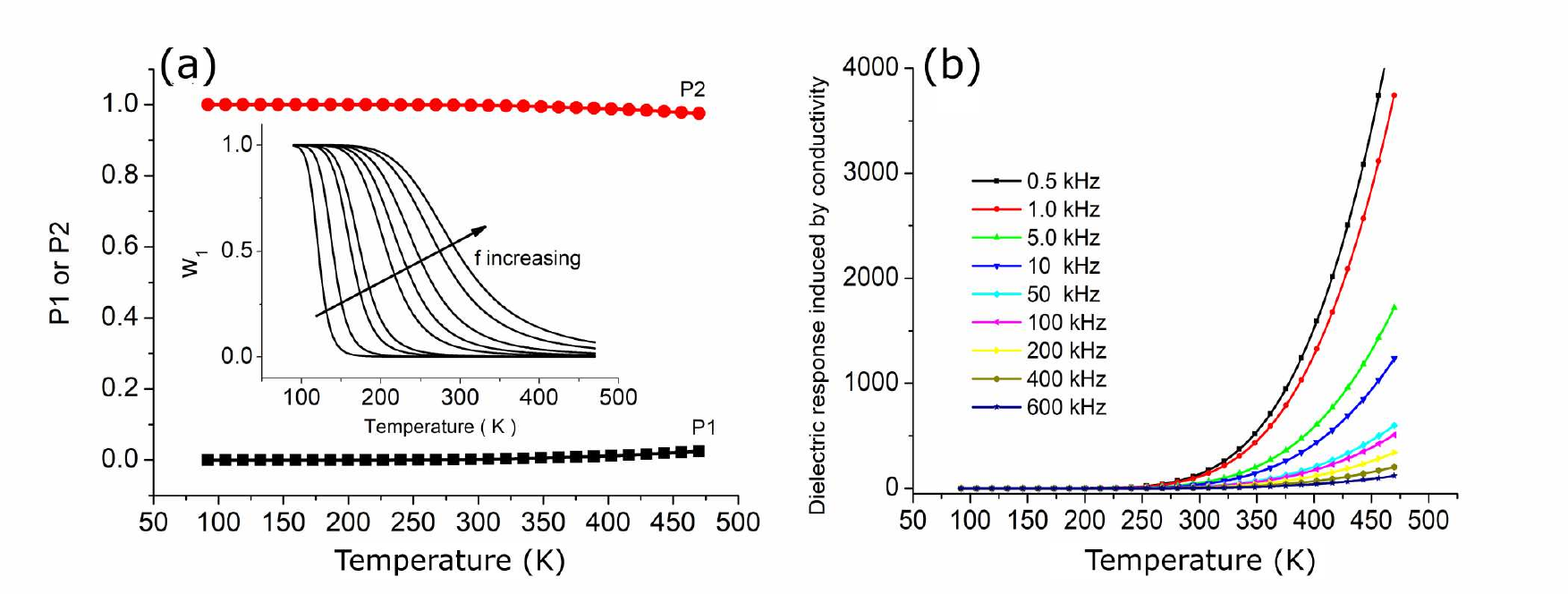}
\par\end{centering}
\caption{(a) The Maxwell-Boltzmann distribution ($P_{1}$ and $P_{2}$) and
the $\varepsilon_{1}/\left[1+\omega^{2}\tau_{0}^{2}\textrm{exp}\left(-\theta/T\right)\right]$
(inset) vs temperature of the CCTO. (b) Temperature dependence on
the dielectric response induced by conductivity.\label{fig3:MBdis_and_Conductivity}}
\end{figure*}
To understand the dielectric response of CCTO, we also show $P_{1}\left(E_{b},T\right)$,
$P_{2}\left(E_{b},T\right)$ and the function $\varepsilon_{1}/\left[1+\omega^{2}\tau_{0}^{2}\textrm{exp}\left(-\theta/T\right)\right]$
in Fig. \ref{fig3:MBdis_and_Conductivity}(a). Clearly, $P_{1}\left(E_{b},T\right)$
and $P_{2}\left(E_{b},T\right)$ show little change as temperature
increases, which is different from typical ferroelectric relaxors,
such as Ba(Ti$_{1-x}$Zr$_{x}$)O$_{3}$ for $x>0.3$ with smaller
$E_{b}$.\cite{LiuLJ3} This feature indicates that the number of
``active'' polarons which can overcome the potential confinement
remain small ($P_{1}$ is small) over the experimental temperature
range. However, the large value of $\varepsilon_{1}$ ($1.25\times10^{6}$
to $8.97\times10^{9}$ depending on the probing frequency, which is
also much larger than that of typical relaxors) indicates that those
polarons are highly correlated and can make an important contribution
to the total permittivity once they overcome $E_{b}$. The function
$\varepsilon_{1}/\left[1+\omega^{2}\tau_{0}^{2}\textrm{exp}\left(-\theta/T\right)\right]$
describes the ability of polarons (which can overcome the potential
well) to align with each other under thermal fluctuations. Figure
\ref{fig3:MBdis_and_Conductivity}(a) shows that it is similar to
the Fermi-Dirac function. That is, at low temperature the value is
close to one but close to zero at high temperatures. We note that
the trailing edges of the curves in the inset of Fig. \ref{fig3:MBdis_and_Conductivity}(a)
correspond to the first rapid increase of the permittivity in Fig.
\ref{fig2:Temperature_vs_ Permittivity}, which shifts to high temperatures
with the increase in frequency. Such characteristics are consistent
with the fact that polarons are susceptible to thermal fluctuations
(and difficult to align with each other), which can easily damage
their dielectric response at higher temperatures.

Figure \ref{fig3:MBdis_and_Conductivity}(b) shows the permittivity
induced by conductivity {[}the 3rd term in Eq. \eqref{eq:total-permittivity}{]}.
The contribution of conductivity is small below $200$\,K but increases
rapidly beyond this point. This is consistent with the basic assumption
of the statistical model that considers thermally activated processes.
At higher temperatures, due to the thermal activation, conductive
charge carriers start to have a large contribution to the dielectric
permittivity. At room temperature, the contribution is about $5\,\%$
of the total dielectric permittivity of CCTO, which suggests that,
at room temperature, polarons make the most important contribution
to the total permittivity. 

\subsection{Polarons and conductive charge carriers}

Zhang \textit{et. al} considered that traces of Ti$^{3+}$ exists
in CCTO due to the loss of oxygen from the grains during the high-temperature
sintering,\cite{ZhangLL} where the Ti$^{3+}$ and Ti$^{4+}$ can
form Ti$^{3+}$-O-Ti$^{4+}$ bonds. The Ti-3\textit{d} electrons in
Ti$^{3+}$ ions can thus hop to Ti$^{4+}$ under an applied electric
field. Moreover, the formation of Ti$^{3+}$ ions distorts CCTO lattices
since the ionic radius of Ti$^{4+}$ is smaller than that of Ti$^{3+}$,
therefore producing a polaronic distortions. The relaxation time of
polarons is larger than that of free electrons because of the polaronic
distortion. 

In the low-temperature region, polarons stay in more localized states
due to the Anderson localization resulting from atom alloying and
strain.\cite{WangCC} Therefore, $E_{b}$ of the CCTO could be high,
comparable to that of ferroelectric Ba(Ti$_{0.9}$Zr$_{0.1}$)O$_{3}$
and Ba(Ti$_{0.8}$Zr$_{0.2}$)O$_{3}$ discussed in Ref. \cite{LiuLJ3}.
In this respect, the dynamic behavior of polarons in CCTO is similar
to that of dipoles of normal ferroelectrics. However, the polarization
changes in CCTO (due to hoppings of polarons) are conceivably much
larger than that of ferroelectrics (due to dipoles on each lattice
site), making the permittivity of CCTO very large. Furthermore, the
hopping of charge carriers (polarons) leads to semiconducting grains,
which in turn leads to the Maxwell-Wagner effect. The herein proposed
model thus helps us to identify the activation energy, and estimate
the relative dielectric strength from a different groups of polarons.
From this perspective, if we want to increase the polaron contribution,
one effective way is to decrease $E_{b}$ so that more polarons can
be excited at room temperature.

It is important to bear in mind that polarons are not the only ones
that can make a substantial contribution to the permittivity of CCTO.
Since CCTO ceramics consist of semiconducting grains and insulating
grain boundaries, the conductivity insides grains increases with temperature
(which eventually make a large contribution to the permittivity at
high temperatures). Therefore, the Maxwell-Wagner effect, which results
in giant dielectric permittivity, starts to be important at the room
temperature (see Fig. \ref{fig1:Temperature_vs_dielectric_loss}).
Further increasing temperature, the diffuse dielectric anomaly is
observed, where the electrical conductivity that is associated with
thermally activated electrons from Ti-3\textit{d} electrons in Ti$^{3+}$
ions (the activation energy $E_{\textrm{con}}$ is much lower than
that of oxygen vacancies) overcome grain boundaries, becoming high
enough to be prominent and need to be included in the permittivity
as shown in Eq. \ref{eq:total-permittivity}. While such effects push
the permittivity of CCTO to even larger values, the accompanying dielectric
loss is so severe that some balance shall be considered to avoid this
situation in the design of other dielectric materials similar to CCTO.

\section{Conclusions}

We have investigated the low frequency 0.5-600\,kHz dielectric properties
of CCTO over a large temperature range ($90-500$\,K). We have proposed
an explicit formula to fit the temperature dependence of the dielectric
permittivity at different frequencies. Our statistical model can explain
well the low temperature step-function-like dielectric relaxation
of CCTO and has estimated the activation energy of polarons. The contribution
from conductive charge carriers at high temperatures also plays a
key role on the dielectric behavior. We hope that our model and the
fitting results will deepen the understanding of the behavior of CCTO
resulting from the interplay between the localization and conduction
of charge carriers. 
\begin{acknowledgement}
This work was supported financially by the National Natural Science
Foundation of China (NSFC), Grant Nos. 11564010, 11574246, 51390472,
and U1537210, and the National Basic Research Program of China, Grant
No. 2015CB654903, and the Natural Science Foundation of Guangxi (Grant
No. GA139008, CB380006, FA198015).
\end{acknowledgement}


\begin{thebibliography}{10}
\bibitem{Subramanian}Subramanian, M. A.; Dong, L.; Duan, N.; Reisner,
B. A.; Sleight, A. W. High Dielectric Constant in \textit{A}Cu$_{3}$Ti$_{4}$
O$_{12}$ and \textit{A}Cu$_{3}$ Ti$_{3}$ FeO$_{12}$ Phases. \textit{J.
Solid State Chem. }\textbf{2000}, \textit{151}, 323-325. 

\bibitem{Ramirez}Ramirez, A. P.; Subramanian, M. A.; Gardel, M.;
Blumberg, G.; Li, D.; Vogt, T.; Shapiro, S. M. Giant Dielectric Constant
Response in a Copper-titanate. \textit{Solid State Commun.} \textbf{2000},
\textit{115}, 217-220. 

\bibitem{Homes}Homes, C. C.; Vogt, T.; Shapiro, S. M.; Wakimoto,
S.; Ramirez, A. P. Optical Response of High-dielectric-constant Perovskite-related
Oxide. \textit{Science} \textbf{2001}, \textit{293}, 673-676. 

\bibitem{Lin}Lin, Y.; Chen, Y. B.; Garret, T.; Liu, S. W.; Chen,
C. L.; Chen, L.; Bontchev, R. P.; Jacobson A.; Jiang, J. C.; Meletis,
E. I.; Horwitz, J.; Wu, H.-D. Epitaxial Growth of Dielectric CaCu$_{3}$Ti$_{4}$O$_{12}$
Thin Films on (001) LaAlO$_{3}$ by Pulsed Laser Deposition. \textit{Appl.
Phys. Lett.} \textbf{2002}, \textit{81}, 631-633. 

\bibitem{Si}Si, W.; Cruz, E. M.; Johnson, P. D.; Barnes, P. W.; Woodward,
P.; Ramirez, A. P. Epitaxial Thin Films of The Giant-dielectric-constant
Material CaCu$_{3}$Ti$_{4}$O$_{12}$ Grown by Pulsed-laser Deposition.\textit{
Appl. Phys. Lett.} \textbf{2002}, \textit{81}, 2056-2058.

\bibitem{Felix}Felix, A. A.; Orlandi, M. O.; Varela, J. A. Schottky-type
Grain Boundaries in CCTO Ceramics. \textit{Solid State Commun.} \textbf{2011},
\textit{151}, 1377-1381.

\bibitem{Schmidt}Schmidt, R.; Stennett, M. C.; Hyatt, N. C.; Pokorny,
J.; Prado-Gonjal, J.; Li, M.; Sinclair, D. C. Effects of Sintering
Temperature on The Internal Barrier Layer Capacitor (IBLC) Structure
in CaCu$_{3}$Ti$_{4}$O$_{12}$ (CCTO) Ceramics.\textit{ J. Eur.
Ceram. Soc. }\textbf{2012}, \textit{32}, 3313-3323.

\bibitem{Han}Han, F.; Ren, S.; Deng, J.; Yan, T.; Ma, X.; Peng, B.;
Liu, L. Dielectric Response Mech\textit{J}anism and Suppressing High-frequency
Dielectric Loss in Y$_{2}$O$_{3}$ Grafted CaCu$_{3}$Ti$_{4}$O$_{12}$
Ceramics.\textit{ J. Mater. Sci.: Mater. Electron.} \textbf{2017},
\textit{28}, 17378-17387.

\bibitem{Deng}Deng, J.; Sun, X.; Liu, L.; Liu, S.; Huang, Y.; Fang,
L.; Elouadi, B. Dielectric Properties of SrMnO$_{3}$-doped K$_{0.5}$Na$_{0.5}$NbO$_{3}$
Lead-free Ceramics. \textit{J. Adv. Dielect.} \textbf{2016}, \textit{6},
No. 1650009.

\bibitem{Deng10}Deng, J.; Liu, L.; Sun, X.; Liu, S.; Yan, T.; Fang,
L.; Elouadi, B. Dielectric Relaxation Behavior and Mechanism of Y$_{2/3}$Cu$_{3}$Ti$_{4}$O$_{12}$
Ceramic. \textit{Mater. Res. Bull.} \textbf{2017}, \textit{88}, 320.

\bibitem{Neupane}Neupane, K. P.; Cohn, J. L.; Terashita, H.; Neumeier,
J. J. Doping Dependence of Polaron Hopping Energies in La$_{1-x}$Ca$_{x}$MnO$_{3}$
$\left(0\leqslant x\leqslant0.15\right)$. \textit{Phys. Rev. B} \textbf{2006},
\textit{74}, No. 144428.

\bibitem{Freitas}Freitas, R. S.; Mitchell, J. F.; Schiffer, P. Magnetodielectric
Consequences of Phase Separation in The Colossal Magnetoresistance
Manganite Pr$_{0.7}$Ca$_{0.3}$MnO$_{3}$. \textit{Phys. Rev. B}
\textbf{2005},\textit{ 72}, No. 144429.

\bibitem{Wang}Wang, C. C.; Cui, Y. M.; Zhang, L. W. Dielectric Properties
of TbMnO$_{3}$ Ceramics. \textit{Appl. Phys. Lett.} \textbf{2007},
\textit{90}, No. 012904.

\bibitem{Yang}Yang, A. M.; Sheng, Y. H.; Farid, M. A.; Zhang, H.;
Lin, X. H.; Li, G. B.; Liu L. J.; Liao F. H.; Lin, J. H. Copper Doped
EuMnO$_{3}$: Synthesis, Structure and Magnetic Properties. \textit{RSC
Adv.} \textbf{2016}, \textit{6}, 13928-13933.

\bibitem{Deng1}Deng, J.; Yang A.; Farid M. A.; Zhang, H.; Li, J.;
Zhang, H.; Li, G.; Liu, L.; Sun, J.; Lin, Synthesis, Structure and
Magnetic Properties of (Eu$_{1-x}$Mn$_{x}$) MnO$_{3-\delta}$. \textit{RSC
Adv.} \textbf{2017}, \textit{7}, 2019-2024.

\bibitem{Deng2}Deng, J.; Farid, M. A.; Yang, A.; Zhang, J.; Zhang,
H.; Zhang, L.; Qiu, Y.; Yu, M.; Zhu, H.; Zhong, M; Li J.; Li, G.;
Liu, L.; Sun, J.; Lin, J. The Origin of Multiple Magnetic and Dielectric
Anomalies of Mn-doped DyMnO$_{3}$ in Low Temperature Region. \textit{J.
Alloys \& Compd.} \textbf{2017}, \textit{725}, 976-983.

\bibitem{LiG}Li, G.; Chen, Z.; Sun, X.; Liu, L.; Fang, L.; Elouadi,
B. Electrical Properties of AC$_{3}$B$_{4}$O$_{12}$-type Perovskite
Ceramics with Different Cation Vacancies. \textit{Mater. Res. Bull.}
\textbf{2015}, 65, 260-265.

\bibitem{ParkT}Park, T.; Nussinov, Z.; Hazzard, K. R. A.; Sidorov,
V. A.; Balatsky, A. V.; Sarrao, J. L.; Cheong, S.-W.; Hundley, M.
F.; Lee, J.-S.; Jia, Q. X.; Thompson, J. D. Novel Dielectric Anomaly
in The Hole-Doped La$_{2}$Cu$_{1-x}$Li$_{x}$O$_{4}$ and La$_{2-x}$Sr$_{x}$NiO$_{4}$
Insulators: Signature of an Electronic Glassy State. \textit{Phys.
Rev. Lett.} \textbf{2005}, \textit{94}, No. 017002.

\bibitem{Rivas}Rivas, J.; Rivas-Murias, B.; Fondado, A.; Mira, J.;
Senaris-Rodriguez, M. A. Dielectric Response of The Charge-ordered
Two-dimensional Nickelate La$_{1.5}$Sr$_{0.5}$NiO$_{4}$. \textit{Appl.
Phys. Lett.} \textbf{2004}, \textit{85}, 6224-6226.

\bibitem{Wu}Wu, J.; Nan, C. W.; Lin, Y.; Deng, Y. Giant Dielectric
Permittivity Observed in Li and Ti Doped NiO. \textit{Phys. Rev. Lett.}
\textbf{2002}, \textit{89}, No. 217601.

\bibitem{LiY}Li, Y.; Fang, L.; Liu, L.; Huang, Y.; Hu, C. Giant Dielectric
Response and Charge Compensation of Li-and Co-doped NiO ceramics.
\textit{Mater. Sci. Eng. B} \textbf{2012}, \textit{177}, 673-677.

\bibitem{Raevski}Raevski, I. P.; Prosandeev, S. A.; Bogatin, A. S.;
Malitskaya, M. A.; Jastrabik, L. High Dielectric Permittivity in AFe$_{1/2}$B$_{1/2}$O$_{3}$
Nonferroelectric Perovskite Ceramics (A= Ba, Sr, Ca; B= Nb, Ta, Sb).
\textit{J. Appl. Phys.} \textbf{2003}, \textit{93}, 4130-4136.

\bibitem{Ke}Ke, S.; Lin, P.; Huang, H.; Fan, H.; Zeng, X. Mean-field
Approach to Dielectric Relaxation in Giant Dielectric Constant Perovskite
Ceramics. \textit{J. Ceram.} \textbf{2013}, \textit{2013}, No. 795827.

\bibitem{Huang}Huang, Y.; Shi, D.; Liu, L.; Li, G.; Zheng, S.; Fang,
L. High-temperature Impedance Spectroscopy of BaFe$_{0.5}$Nb$_{0.5}$O$_{3}$
Ceramics Doped with Bi$_{0.5}$Na$_{0.5}$TiO$_{3}$. \textit{Appl.
Phys. A} \textbf{2014}, \textit{114}, 891-896.

\bibitem{Lius}Liu, S.; Sun, X.; Peng, B.; Su, H.; Mei, Z.; Huang,
Y.; Deng, J.; Su, C.; Fang, L.; Liu, L. Dielectric Properties and
Defect Mechanisms of $(1-x)$Ba(Fe$_{0.5}$Nb$_{0.5}$)O$_{3}$-$x$BiYbO$_{3}$
Ceramics. \textit{J. Electroceram.} \textbf{2016}, \textit{37}, 137-144.

\bibitem{SunX}Sun, X.; Deng, J.; Liu, S.; Yan, T.; Peng, B.; Jia,
W.; Mei, Z.; Su, H.; Fang, L.; Liu, L. Grain Boundary Defect Compensation
in Ti-doped BaFe$_{0.5}$Nb$_{0.5}$O$_{3}$ Ceramics. \textit{Appl.
Phys. A} \textbf{2016}, \textit{122}, No. 864.

\bibitem{WangCC}Wang, C. C.; Zhang, L. W. Polaron Relaxation Related
to Localized Charge Carriers in CaCu$_{3}$Ti$_{4}$O$_{12}$. \textit{Appl.
Phys. Lett.} \textbf{2007}, \textit{90}, No. 142905.

\bibitem{ZhangLL}Zhang, L.; Tang, Z. J. Polaron relaxation and variable-range-hopping
conductivity in the giant-dielectric-constant material CaCu$_{3}$Ti$_{4}$O$_{12}$.
\textit{Phys. Rev. B} \textbf{2004}, \textit{70}, No. 174306.

\bibitem{Tselev}Tselev, A.; Brooks, C. M.; Anlage, S. M.; Zheng,
H.; Salamanca-Riba, L.; Ramesh, R.; Subramanian, M. A. (2004). Evidence
for Power-law Frequency Dependence of Intrinsic Dielectric Response
in The CaCu$_{3}$Ti$_{4}$O$_{12}$. \textit{Phys. Rev. B }\textbf{2004},
\textit{70}, No. 144101.

\bibitem{Valdez-Nava}Valdez-Nava, Z.; Guillemet-Fritsch, S.; Tenailleau,
C.; Lebey, T.; Durand, B.; Chane-Ching, J. Y. Colossal Dielectric
Permittivity of BaTiO$_{3}$-based Nanocrystalline Ceramics Sintered
by Spark Plasma Sintering. \textit{J. Electroceramics} \textbf{2009},
\textit{22}, 238-244.

\bibitem{LiuLJ}Liu, L.; Fan, H.; Wang, L.; Chen, X.; Fang, P. Dc-bias-field-induced
Dielectric Relaxation and ac Conduction in CaCu$_{3}$Ti$_{4}$O$_{12}$
Ceramics. \textit{Philos. Mag.} \textbf{2008}, \textit{88}, 537-545.

\bibitem{LiuLJ2}Liu, L.; Shi, D.; Zheng, S.; Huang, Y.; Wu, S.; Li,
Y.; Fang, L.; Hu, C. Polaron Relaxation and Non-ohmic Behavior in
CaCu$_{3}$Ti$_{4}$O$_{12}$ Ceramics with Different Cooling Methods.
\textit{Mater. Chem. Phys.} \textbf{2013}, \textit{139}, 844-850.

\bibitem{Bidault}Bidault, O.; Maglione, M.; Actis, M.; Kchikech,
M.; Salce, B. Polaronic Relaxation in Perovskites. \textit{Phys. Rev.
B} \textbf{1995}, \textit{52}, 4191-4197.

\bibitem{Mott}Mott, N. F.; Davis, E. A. \textit{Electronic Processes
in Non-crystalline Solids}. Clarendon: Oxford , 1979.

\bibitem{Krohns}Krohns, S.; Lunkenheimer, P.; Ebbinghaus, S. G.;
Loidl, A. Colossal Dielectric Constants in Single-crystalline and
Ceramic CaCu$_{3}$Ti$_{4}$O$_{12}$ Investigated by Broadband Dielectric
Spectroscopy. \textit{J. Appl. Phys}. \textbf{2008}, \textit{103},
084107.

\bibitem{Chung}Chung, S. Y.; Kim, I. D.; Kang, S. J. L. Strong Nonlinear
Current\textendash Voltage Behaviour in Perovskite-derivative Calcium
Copper Titanate. \textit{Nature Mater.} \textbf{2004}, \textit{3},
774-778.

\bibitem{Kastner}Kastner, M. A.; Birgeneau, R. J.; Chen, C. Y.; Chiang,
Y. M.; Gabbe, D. R.; Jenssen, H. P.; Junk, T.; Peter, C. J.; Picone,
P. J.; Thio, T.; Thurston, T. R.; Tuller, H. L. Resistivity of Nonmetallic
La$_{2-y}$Sr$_{y}$Cu$_{1-x}$Li$_{x}$O$_{4-\delta}$ Single Crystals
and Ceramics. \textit{Phys. Rev. B} \textbf{1988}, \textit{37}, 111-117.

\bibitem{Ang}Ang, C.; Jing, Z.; Yu, Z. Variable-range-hopping Conduction
and Metal-insulator Transition in Cu-doped BaTiO$_{3}$. \textit{J.
Phys.: Condens. Matter }\textbf{1999}, \textit{11}, 9703-9708.

\bibitem{Karmakar}Karmakar, A.; Majumdar, S.; Giri, S. Polaron Relaxation
and Hopping Conductivity in LaMn$_{1-x}$Fe$_{x}$O$_{3}$. \textit{Phys.
Rev. B} \textbf{2009}, \textit{79}, No. 094406.

\bibitem{Liuj}Liu, J.; Li, F.; Zeng, Y.; Jiang, Z.; Liu, L.; Wang,
D.; Ye, Z.-G.; Jia, C. L. Insights into The Dielectric Response of
Ferroelectric Relaxors from Statistical Modeling. \textit{Phys. Rev.
B} \textbf{2017}, \textit{96}, No. 054115.

\bibitem{Raevski2003a}Raevski, I. P.; Prosandeev, S. A.; Bogatina,
A. S.; Malitskaya, M. A., and Jastrabik L. High-k ceramic materials
based on nonferroelectric AFe$_{1/2}$B$_{1/2}$O$_{3}$ (A-Ba, Sr,
Ca; B-Nb, Ta, Sb) perovskites. Integrated Ferroelectrics \textbf{2003},
55, 757-768.

\bibitem{Raevski2003b} Raevski, I. P.; Prosandeev, S. A.; Bogatin,
A. S.; Malitskaya, M. A., and Jastrabik L. J. Appl. Phys. \textbf{2003},
93, 4130. 

\bibitem{Maglione2016} Maglione M., Free charge localization and
effective dielectric permittivity in oxides, J. Adv, Delect. 2016,
\textbf{6}, 163006.

\bibitem{Kasap}Kasap, S. O. \textit{Principles of Electronic Materials
and Devices}, Third ed; McGraw-Hill: New York, 2006.

\bibitem{Jonscher}Jonscher, A. K. A New Understanding of The Dielectric
Relaxation of Solids. \textit{J. Mater. Sci.} \textbf{1981}, \textit{16},
2037-2060.

\bibitem{LiuLJ3}Liu, L.; Ren, S.; Zhang, J.; Peng, B.; Fang, L.;
Wang, D. Revisiting The Temperature-dependent Dielectric Permittivity
of Ba(Ti$_{1-x}$Zr$_{x}$)O\textit{$_{3}$. J. Am. Ceram. Soc.} \textbf{2018},
\textit{101}, 2408-2416.
\end{thebibliography}
\end{document}